\documentclass[12pt]{article} 
\usepackage{amsmath}
\begin{document}

\title{Is my ODE a Painlev\'e equation in disguise?}
\author{Jarmo~HIETARINTA~$^\dag$ and Valery~DRYUMA~$^\ddag$\\
$^\dag$ Department of Physics, University of Turku,\\ FIN-20014 Turku,
  Finland, e-mail:~jarmo.hietarinta@utu.fi\\
  $^\ddag$ Institute of Mathematics AS RM, Academy str. 5,\\ 277028 Kishinev,
  Moldova, e-mail:~valery@gala.moldova.su}

\maketitle
\begin{abstract}
\noindent
Painlev\'e equations belong to the class $y'' + a_1\, {y'}^3 + 3 a_2\, {y'}^2
+ 3 a_3\, y' + a_4 = 0$, where $a_i=a_i(x,y)$. This class of equations is
invariant under the general point transformation $x=\Phi(X,Y),\, y=\Psi(X,Y)$
and it is therefore very difficult to find out whether two equations in this
class are related.  We describe R.~Liouville's theory of invariants that can
be used to construct invariant characteristic expressions (syzygies), and in
particular present such a characterization for Painlev\'e equations I-IV.
\end{abstract}

\section{Introduction}
Many phenomena in Nature are modeled by ordinary differential equations
(ODEs). After such an equation is derived for some physical situation, the
natural question is whether that ODE is well known, or at least transformable
to a well known equation.  For example, one would like to know if the equation
is related to a known integrable equation, e.g, in the case of a second order
ODE, to one of the Painlev\'e equations, or to one of the equations in the
Gambier\cite{G}/Ince\cite{I} list.

Normally these lists contain only {\em representative} equations, e.g., up to
some group of transformations. If the equation is of the form $y''=f(x,y,y')$
the usually considered group of transformations is
\[
x=\phi(X),\quad y=\frac{\psi_1(X)Y+\psi_2(X)}{\psi_3(X)Y+\psi_4(X)}.
\]
However, when we have an equation derived from some physical problem the
required transformation may be more complicated.

Here we consider the following class of equations
\begin{equation}
y''+a_1(x,y)\,{y'}^3+3a_2(x,y)\,{y'}^2+3a_3(x,y)\,{y'}+a_4(x,y)=0.
\label{class}
\end{equation}
This class is invariant under a general point transformation
\begin{equation}
x=\Phi(X,Y),\quad y=\Psi(X,Y),
\label{tra}
\end{equation}
we only need to assume that the transformation is nonsingular, i.e.,
\begin{equation}
\Delta:=\Phi_X\Psi_Y-\Phi_Y\Psi_X \neq 0.
\end{equation}
The point transformation (\ref{tra}) prolongs to
\begin{eqnarray}
y'&=&\frac{\Psi_X+\Psi_Y\,Y'}{\Phi_X+\Phi_Y\,Y'},\\
y''&=&\bigl\{[\Psi_Y\Phi_X-\Psi_X\Phi_Y]Y''
+[\Phi_Y\Psi_{YY}-\Psi_Y\Phi_{YY}]{Y'}^3\bigr.\nonumber\\
&&+(\Phi_X \Psi_{YY} - 2 \Phi_{XY} \Psi_Y + 2 \Phi_Y \Psi_{XY}
 - \Phi_{YY} \Psi_X){Y'}^2\nonumber\\ && +( - \Phi_{XX} 
\Psi_Y - 2 \Phi_{XY} \Psi_X +2 \Phi_X \Psi_{XY}+ \Phi_Y \Psi_{XX})Y'
\nonumber\\ &&\bigl. +\Phi_X \Psi_{XX} - \Phi_{XX} \Psi_X
\bigr\}/[\Phi_X+\Phi_Y\,Y']^3,
\end{eqnarray}
and when these are substituted into (\ref{class}) the form of the equation
(cubic in $y'$) stays the same, but the coefficients $a_i$ change as follows:
\begin{eqnarray}
 \widetilde{a_1} &=& \tfrac1{\Delta}( \Phi_Y \Psi_{YY} - \Phi_{YY} \Psi_Y +
\Phi_Y^3 A_4 + 3 \Phi_Y^2 \Psi_Y A_3 + 3 \Phi_Y \Psi_Y^2 A_2 + \Psi_Y^3 A_1),
\label{atra1}\\
\widetilde{a_2} &=& \tfrac1{\Delta}[\tfrac13( \Phi_X \Psi_{YY}
 - 2 \Phi_{XY} \Psi_Y + 2
\Phi_Y \Psi_{XY} - \Phi_{YY} \Psi_X) +  \Phi_X \Phi_Y^2 A_4 \nonumber\\
&&+ (2 \Phi_X \Phi_Y \Psi_Y  + \Phi_Y^2 \Psi_X)  A_3
+ (\Phi_X \Psi_Y^2  + 2 \Phi_Y \Psi_X \Psi_Y)  A_2 +  \Psi_X \Psi_Y^2 A_1],
\label{atra2}\\
\widetilde{a_3} &=& \tfrac1{\Delta}[\tfrac13( - \Phi_{XX} \Psi_Y
 - 2 \Phi_{XY} \Psi_X +
2 \Phi_X \Psi_{XY} + \Phi_Y \Psi_{XX}) +  \Phi_X^2 \Phi_Y A_4 \nonumber\\
&&+ (\Phi_X^2 \Psi_Y  + 2 \Phi_X \Phi_Y \Psi_X)  A_3
 + (2 \Phi_X \Psi_X \Psi_Y  +\Phi_Y \Psi_X^2)  A_2 +  \Psi_X^2 \Psi_Y A_1],
\label{atra3}\\
\widetilde{a_4} &=& \tfrac1{\Delta}( - \Phi_{XX} \Psi_X + \Phi_X \Psi_{XX} +
\Phi_X^3 A_4 + 3\Phi_X^2 \Psi_X A_3 + 3\Phi_X \Psi_X^2 A_2 + \Psi_X^3 A_1).
\label{atra4}
\end{eqnarray}
Here $A_i(X,Y):=a_i(\Phi(X,Y),\Psi(X,Y))$.  Example: If we apply the
transformation $x=X+Y,\, y=XY$ to the equation $y''=0$ (in which $a_i=0$),
then we find $\widetilde{a_1}=\widetilde{a_4}=0$,
$\widetilde{a_2}=\widetilde{a_3}= \tfrac23\frac1{X-Y}$, i.e., the equation
becomes $ Y''+\frac2{X-Y}({Y'}^2+Y')=0$.

Since the coefficients of equation (\ref{class}) transform in such a
complicated way it is in general difficult to find a characterization of
(\ref{class}) that is invariant under a general point transformation
(\ref{tra}).  This classical problem was basically solved more than 100 years
ago, the fundamental works being those by Liouville \cite{L} in 1889, Tresse
\cite{Tre} in 1894, with more modern formulations by Cartan \cite{Car} in 1924
and Thomsen \cite{Tho} in 1930.  Recent wave of interest on this classical
problem started with \cite{Dry}. [For the restricted problem with $x=\Phi(X)$
see, e.g., \cite{Kam}.]

\section{Relative invariants, absolute invariants and syzygies}
The invariants we are looking for must be constructed from the coefficients
$a_i$ and their various derivatives. The transformation rules for $a_i$ were
given in (\ref{atra1}-\ref{atra4}) and we are now looking for some
combinations that transform in a much simpler way.  Let us consider some
expression
\[
I[x,y]=I(a_1,\dots,a_4,\partial_x a_1,\dots,\partial_x a_1,
\partial_y a_1,\dots,\partial_y a_4,\partial_x^2 a_1,\dots).
\]
Since $a_i=a_i(x,y)$ this will be a function of $x,y$.  Under the
transformation (\ref{tra}) the ingredients $a_i$ transform to $\widetilde a_i$
and the expression constructed from the transformed quantities in exactly the
same way is
\[
\widetilde I[X,Y]=I(\widetilde a_1,\dots,\widetilde a_4,
\partial_X \widetilde a_1,\dots,\partial_X \widetilde a_1,
\partial_Y \widetilde a_1,\dots,\partial_Y \widetilde a_4,
\partial_X^2 \widetilde a_1,\dots).
\]
This is now a function of $X,Y$ and if it turns out that
\[
\widetilde I[X,Y]=\Delta^n I[\Phi(X,Y),\Psi(X,Y)],
\]
i.e., that $I$ transforms, up to some overall factor, as by {\em substitution}
then we say that $I$ is a {\bf relative invariant} of weight $n$. If
furthermore $n=0$ or $I=0$ we say that $I$ is an {\bf absolute invariant}.
Normally the weight of the relative invariant is indicated by a subscript.
Later we will construct several sequences of relative invariants, e.g.,
$i_{2n},\,n=1,2,3,\dots$, then using them we can construct a sequence of
absolute invariants, in this case $j_{2n}:=i_{2n}/i_2^n$.

Although the absolute invariants transform by a simple substitution under
(\ref{tra}) they are normally complicated functions of $x,y$ and therefore in
practice useless for classification. What we need are {\em relationships
  between absolute invariants}, i.e., {\bf syzygies}.

For example, it turns out later that for one particular equation
$j_6-6j_4+4=0$.  This relationship is invariant under the transformation
(\ref{tra}) and therefore it is an invariant characterization for that
equation.  Using this result we can say with certainty that any equation that
does {\em not} satisfy this relationship cannot be transformed to first
equation by {\em any} point transformation.  The reverse is not true: several
equations may satisfy the same syzygy.

Note also that the syzygy polynomials can only have numerical coefficients,
because parameter values appearing in the equation can be changed with the
allowed transformations.

\section{Construction of Invariants}
The construction of the relative and absolute invariants proceeds step by step
as follows: First define
\begin{eqnarray}
\Pi_{22}&=&\partial_x a_1-\partial_y a_2+2(a_2^2-a_1a_3),\\
\Pi_{12}&=&\partial_x a_2-\partial_y a_3+a_2a_3-a_1a_4,\\
\Pi_{11}&=&\partial_x a_3-\partial_y a_4+2(a_3^2-a_2a_4),
\end{eqnarray}
and then
\begin{eqnarray}
  L_2&=&\partial_x \Pi_{22}-\partial_y\Pi_{12}-a_1\Pi_{11}+2a_2\Pi_{12}
-a_3\Pi_{22},\\
  L_1&=&\partial_x \Pi_{12}-\partial_y\Pi_{11}-a_2\Pi_{11}+2a_3\Pi_{12}
-a_4\Pi_{22}.
\end{eqnarray}
Now one finds that the $L_i$ transform according to
\begin{equation}
\pmatrix \widetilde{L_1} \\ \widetilde{L_2}  \endpmatrix
=\Delta\,\pmatrix \Phi_X & \Psi_X \\ \Phi_Y & \Psi_Y \endpmatrix
\pmatrix {L_1}[\Phi(X,Y),\Psi(X,Y)] \\ {L_2}[\Phi(X,Y),\Psi(X,Y)]
\endpmatrix,
\end{equation}
which is already a relatively simple transformation rule.
From this we get the first result (R. Liouville):
\vskip 0.3cm 

\noindent{\em %\Ovalbox{\begin{minipage}{13cm}
The property $L_1=L_2=0$ is an absolute invariant 
and if it holds the  equation can be transformed to $Y''=0$.
}%\end{minipage}}
\vskip 0.3cm

If we define $a_i$ and $\partial_x,\,\partial_y$ to have weight $\frac12$,
then $L_i$ are of weight $\frac32$. Continuing with $L_i$ and adding one
derivative or $a_i$ we find at weight $\frac72$ another pair
\begin{eqnarray}
\hskip -0.7cm 
Z_1&=&-L_{1y}L_1-3L_{1x}L_2+4L_1L_{2x}+3a_2L_1^2-6a_3L_1L_2+3a_4L_2^2,\\
\hskip -0.7cm
Z_2&=&\phantom{-}L_{2x}L_2+3L_{2y}L_1-4L_2L_{1y}+3a_1L_1^2-6a_2L_1L_2+
3a_3L_2^2 ,
\end{eqnarray}
that transforms similarly:
\begin{equation}
\pmatrix \widetilde{Z_1} \\ \widetilde{Z_2}  \endpmatrix
=\Delta^3\,\pmatrix \Phi_X & \Psi_X \\ \Phi_Y & \Psi_Y \endpmatrix
\pmatrix {Z_1}[\Phi(X,Y),\Psi(X,Y)] \\ {Z_2}[\Phi(X,Y),\Psi(X,Y)]   \endpmatrix
\end{equation}
Using these two we get the first semi-invariant
\begin{eqnarray}
  \nu_5&=&\tfrac13[Z_1L_2-Z_2L_1]\nonumber\\
&=&L_2(L_1\partial_xL_2-L_2\partial_xL_1)+
  L_1(L_2\partial_yL_1-L_1\partial_yL_2)\nonumber\\
  &&-a_1L_1^3+3a_2L_1^2L_2-3a_3L_1L_2^2+a_4L_2^3,
 \label{eq:nu5}
\end{eqnarray}
which is of weight 5, i.e., transforms as $\widetilde{\nu_5}=
\Delta^5\,\nu_5$.

\vskip 0.3cm
Observation: {\em For all Painlev\'e equations $\nu_5=0$.}

\vskip 0.3cm Our ultimate aim is to provide an invariant characterization for
the Gambier/Ince list of 50 equations. It contains some equations with
$L_1=L_2=0$, and other equations besides Painlev\'e's that have $\nu_5=0$, but
it also contains many with $\nu_5\neq0$.  [We note in passing, that in the
standard form all Painlev\'e equations also have the properties $a_1=0$ and
$L_2=0$, but these are not (semi)invariant characterizations.]

Let us now continue with $\nu_5=0$, i.e. $Z_1L_2=Z_2L_1$. Let us define
\begin{eqnarray}
R_1&=&L_1\partial_xL_2-L_2\partial_xL_1+a_2L_1^2-2a_3L_1L_2+a_4L_2^2,\\
  w_1&=&[L_1^3(\Pi
_{11}L_2-\Pi_{12}L_1)+R_1\partial_x(L_1^2)
-L_1^2\partial_xR_1%\nonumber\\ && \quad 
+L_1R_1(a_3L_1-a_4L_2)]/L_1^4.
  \label{eq:w1}
\end{eqnarray}
[If $L_1=0$ there is a similar expression with $L_2$ as divider.] The
expression $w_1$ is a semi-invariant of weight 1.

\vskip 0.3cm
Observation: {\em For all Painlev\'e equations $w_1=0$.}

\vskip 0.3cm \noindent The Gambier/Ince list contains also equations with
$\nu_5=0,\,w_1\neq 0$.

We continue further with $\nu_5=0,\,w_1=0$. A sequence of semi-invariants can
now be constructed starting with
\begin{equation}
  \label{eq:i2}
  i_2=2R_1/L_1+\partial_xL_2-\partial_yL_1,
\end{equation}
with higher members given recursively by
\begin{equation}
  \label{eq:im}
  i_{2(m+1)}=L_1\,\partial_y i_{2m}-L_2\,\partial_x i_{2m}
+2m\,i_{2m}(\partial_x L_2-\partial_y L_1).
\end{equation}
If $i_2\neq 0$ a sequence of absolute invariants is given by
\begin{equation}
  \label{eq:jm}
  j_{2m}=i_{2m} i_2^{-m}.
\end{equation}
In the next section we will use the $j_{2m}$ to characterize Painlev\'e
equations I-IV.

At this point we can see that the classification using $j_{2m}$ cannot be
sharp. Consider the special case $y''+a_4(x,y)=0$. Then
$
\Pi_{11}=-\partial_y a_4,\,\Pi_{12}=\Pi_{22}=0, L_1=\partial_y^2 a_4, L_2=0,
\nu_5=0,\,w_1=0,
$
and
\[
i_2=-\partial_y^3 a_4,\quad  i_{2(m+1)}=L_1^{2m+1}\partial_y(i_{2m}/L_1^{2m}),
\]
The semi-invariants therefore depend only on $ \partial_y^2 a_4$
and its higher $y$-derivatives and are insensitive to the possible linear in
$y$ part in $a_4$.

\section{Invariant characterization of $P_I-P_{IV}$}
The first steps were given above: All Painlev\'e equations have the properties
1) at least one of $L_1,L_2$ is nonzero, 2) $\nu_5=0$, 3)
$w_1=0$.\footnote{This was first observed by V.~Dryuma in late 1980's}  If
the candidate equation fails any of these properties it cannot be transformed
to a Painlev\'e equation using the transformation (\ref{tra}).  Here we want
to go further and derive conditions which differentiate between Painlev\'e
equations.  The following results have been derived using the symbolic algebra
language REDUCE\cite{RE}.  It should be noted that we give here only the lowest
degree syzygy.

\subsection{Painlev\'e I}
Any equation of the form 
\begin{equation}
y''=6y^2+f_1(x)y+f_0(x),\label{P1}
\end{equation}
has the property $i_2=0$. In principle $i_2$ is only a relative invariant, but
since its value in this case is 0, this property is an absolute invariant.
Equation (\ref{P1}) contains Painlev\'e I for which $f_1=0$, $f_0=x$. For all
other Painlev\'e equations $i_2\neq 0$.

\subsection{Painlev\'e II}
All equations of form
\begin{equation}
y''=2y^3+f_1(x)y+f_0(x),
\label{P2}
\end{equation}
have the property $i_2=12,i_4=288$ and therefore $j_4=2$ is the syzygy for
this class of equations. [In fact it is easy to see that $j_{2(m+1)}=2^m\,
m!$] Painlev\'e II is contained as the special case $f_1=x,\,f_0=$const.

\subsection{Painlev\'e III} 
The following results are valid for Cases 12,13,$13_1$ in the Ince/Gambier
classification, they all have four parameters, $\alpha,\beta,\gamma,\delta$:
\begin{eqnarray*}
&&y''-\frac{{y'}^2}y-\gamma y^3-\alpha y^2-\beta-\delta/y=0,\\
&&y''-\frac{{y'}^2}y+\frac{y'}x-\gamma y^3-\frac1x (\alpha y^2+\beta)-\delta/y=0,\\
&&y''-\frac{{y'}^2}y-e^x (\alpha y^2+\beta)-e^{2 x} (\gamma y^3+\delta/y)=0.
\end{eqnarray*}
\begin{itemize}
\item[1:] If any 3 of the parameters $\alpha,\beta,\gamma,\delta$ are zero,
  then $j_{2(m+1)}=m!$
  
\item[2a:] If $\gamma=\delta=0$ or $\alpha=\beta=0$ then
\begin{equation}
j_6-6j_4+4=0.
\end{equation}
These two cases are connected: $x=\frac12X^2,y=Y^2$ takes
$(\alpha,\beta,0,0)\to (0,0,\alpha,\beta)$.
\item[2b:] If $\alpha=\gamma=0$ or $\beta=\delta=0$ then\[2 j_8-13   j_6-22
   +57 j_4  -21 j_4^2=0.\] Transformation $x=X,y=1/Y$ takes
  $(0,\beta,0,\delta)\to(-\beta,0,-\delta,0)$.
\item[2c:] If $\beta=\gamma=0$ or $\alpha=\delta=0$ then\[2 j_8-23   j_6-42
   +87 j_4  -11 j_4^2=0.\] Transformation $x=X,y=1/Y$ takes
  $(\alpha,0,0,\delta)\to(0,-\alpha,-\delta,0)$.
\item[3:] If $\alpha^2\delta+\beta^2\gamma=0$ (which contains 2a,2b)
  then we get a degree 12 relation
\begin{eqnarray*}
 && 4 j_{12} - 76 j_{10}   - 1696   + 6840   j_4 - 2600   j_6 -
  5640   j_4^2\\ 
&&\qquad + 590   j_8 + 2220   j_4 j_6 + 450 j_4^3 - 155 j_4
  j_8 - 115 j_6^2=0.\end{eqnarray*}
\item[4:] For the generic case we obtain
\begin{eqnarray*}
\hskip -2cm &&4 j_{14} - 112 j_{12}   - 231 j_{10} j_4 + 1274
  j_{10}   - 385 j_8 j_6 + 4795 j_8 j_4   - 7910 j_8   + 3255 j_6^2
\\\hskip -2cm  &&\qquad    + 3570 j_6 j_4^2 - 39060 j_6 j_4   + 
30240 j_6   - 15330 j_4^3
    + 78120 j_4^2   - 71736 j_4   + 15264  =0.
\end{eqnarray*}
\end{itemize}

\subsection{Painlev\'e IV} 
Let us again consider a more general form 
\[
y''=\frac1{2y}{y'}^2 +e_1 \frac32 y^3+4 (e_2 x+e_3) y^2+2 (e_4 x^2+e_5 x+e_6)
y+e_7\frac1y,
\]
Painlev\'e IV corresponds to $e_1=e_2=e_4=1,\, e_3=e_5=0,\,
e_6=-\alpha,\, e_7=-\beta^2/2.$ The classification is sensitive only to
$e_1,e_2,e_3,e_7$. 
\begin{itemize}
\item[1:] If $e_1=e_2=e_3=e_7=0$ then $L_1=L_2=0$
\item[2a:] If $e_2=e_3=e_7=0$ then
$j_{2(m+1)}=\left(\tfrac43\right)^{m}m!$
\item[2b:] If $e_2=e_3=e_1=0$ then
$j_{2(m+1)}=\left(\tfrac45\right)^{m}m!$
\item[2c:] If $e_1=e_7=0$ then  $j_{2(m+1)}=2^{m}m!$
\item[3a:] If $e_7=0$ then \[j_8-11   j_6-24  +46 j_4  -7 j_4^2=0.\]
\item[3b:] If $e_1=0$ then  \[5 j_8-95   j_6-136  +286 j_4  +21 j_4^2=0.\]
\item[3c:] if $e_2=e_3=0$ then \[8125 j_{10}-2165 j_8  -2495 j_6 j_4+13096 j_6
 +12872 j_4^2   -36160 j_4  +11904  =0.\]
\item[4:] Generic case:
\begin{eqnarray*}
&&\hskip -2cm  3 j_{14} + 1955 j_{10}   - 277 j_{10} j_4 - 120 j_{12}   + 31488
    - 131008   j_4 + 57952   j_6 + 103600   j_4^2\\ 
&&\hskip -1cm - 15062
    j_8 - 40486   j_4 j_6 - 18772   j_4^3 + 5236   j_4 j_8 + 990
    j_6^2 + 3775 j_4^2 j_6 - 75 j_6 j_8=0.
\end{eqnarray*}
\end{itemize}

\section{Conclusions}
We have presented here the first results of a project aiming to derive
syzygies for every equation in the Gambier/Ince list.  The biggest problem in
finding the characteristic expressions is that we have to solve huge sets of
rather complicated nonlinear algebraic equations. For Painlev\'e~V the generic
expression is of degree higher than 26 (which is as far as we checked) and for
Painlev\'e VI presumably still much higher.  When this classification is
eventually finished:
\begin{itemize}
\item We get an {\em algorithmic} method for finding out if a given
  equation of type (\ref{class}) has any change of being transformed into one
  of the equations in the Gambier/Ince list.
\item We also get a classification of the equations into essentially different
  sub-cases (c.f. \cite{Ram,Cos}).
\end{itemize}

The first results that have already been obtained suggest some interesting open
questions:
\begin{itemize}
\item Is the type of ``integrability'' different in the classes a)
  $\nu_5=w_1=0$, b) $\nu_5=0,\,w_1\neq0$, and c) $\nu_5\neq0$?
\item What does the degree of the minimal syzygy tell us about integrability?
\end{itemize}

\section*{Acknowledgment}
J.H. would like to thank P. Olver for discussions.  V.D. would like to thank
the Physics Department of University of Turku and INTAS-Program 99-01782
for financial support and kind hospitality.

\label{hietarinta-lastpage}
\label{lastpage}
\end{document}